\documentclass[journal,onecolumn]{IEEEtran}
\usepackage{authblk}
\usepackage[utf8]{inputenc}
\usepackage{lscape,graphicx,epsfig}
\usepackage{amssymb,amsmath,amsfonts,bm,color}
\usepackage{cite}
\usepackage{hyperref}
\usepackage[ruled]{algorithm2e}
\usepackage[noend]{algpseudocode}

\DeclareMathOperator*{\argmin}{argmin }

\DeclareMathOperator{\sgn}{sgn}

\begin{document}

\title{Ordinal UNLOC: Target Localization with Noisy and Incomplete Distance Measures}

\author[1,3]{Mahesh K. Banavar}
\author[1,2]{Shandeepa Wickramasinghe}
\author[1,3]{Monalisa Achalla}
\author[1,2]{Jie Sun}
\affil[ ]{\thanks{This manuscript has been accepted by the IEEE Internet of Things Journal.}}
\affil[1]{Clarkson Center for Complex Systems Science ($C^3S^2$), Clarkson University}
\affil[2]{Department of Mathematics, Clarkson University}
\affil[3]{Department of Electrical \& Computer Engineering, Clarkson University}

\maketitle

\begin{abstract}
A main challenge in target localization arises from the lack of reliable distance measures. This issue is especially pronounced in indoor settings due to the presence of walls, floors, furniture, and other dynamically changing conditions such as the movement of people and goods, varying temperature and air flows. Here, we develop a new computational framework to estimate the location of a target without the need for reliable distance measures. The method, which we term Ordinal UNLOC, uses only ordinal data obtained from comparing the signal strength from anchor pairs at known locations to the target. Our estimation technique utilizes rank aggregation, function learning as well as proximity-based unfolding optimization. As a result, it yields accurate target localization for common transmission models with unknown parameters and noisy observations that are reminiscent of practical settings. Our results are validated by both numerical simulations and hardware experiments. 
\end{abstract}


\section{Introduction}
\label{sec:introduction}
A recent FCC document \cite{FC14, FC15} highlights the need for increasing the accuracy of localization when users call from mobile devices from indoor environments. While the FCC recommends accurate localization to within 3 meters within 30 seconds, in over 90 percent of the test calls the localization error is greater than 100 meters \cite{CS13}. With a majority of 911 calls now being made from wireless devices \cite{CR11}, over 56 percent of which originate from {\it indoor} locations \cite{CC13}, the need for accurate indoor localization is paramount. A poor localization result can cause incorrect estimation of the room, the floor or even the actual building from which the call originates.
In practice, despite being an important problem whose satisfactory solution can greatly improve safety and user experience; and is critical in emergency situations such as rescuing people from collapsed and/or burning buildings, accurate indoor localization remains an open problem. In most buildings, including the majority of schools, hospitals and shopping malls, there is generally no existing infrastructure (such as an ``indoor GPS'') to directly enable accurate localization. While in most cases, it is in principle possible to set up such an infrastructure with WiFi routers or Bluetooth beacons \cite{Tragas07, Matic10, chintalapudi10, faragher15, rida15, Diwate2015, chai16, kriz16}, the indoor setting itself poses a challenge. This is because indoor environments feature rich scattering and complex multipath due to the presence of physical obstacles and barriers (e.g., walls, furniture, and people) and other factors that make reliable measurements of distance unreliable and accurate localization seemingly hopeless.

Here, we propose to tackle the target localization problem from a new perspective. Instead of using the distance measured between anchors and the target, we use comparative signal strengths from anchors {\it pairs} to the target. Such ordinal data can typically be reliably obtained by comparing time-of-arrival (TOA) or received signal strength indicators (RSSI) instead of distances themselves, and do not require accurate account of unknown environmental parameters or noise in the transmission models. Thus, the crucial difference between the proposed approach and conventional localization methods is that in our Ordinal UNLOC approach, accurate distance measures are not required. Instead, all that is needed is to determine, between each given pair of sensors, which one is closer to a given reference sensor. Based on such ordinal data, we estimate the target location by a series of steps that utilize techniques such as rank aggregation, function learning, and unfolding optimization. We show that our approach generally produces accurate target localization across several transmission models under unknown and changing parameters as well as measurement noise, a scenario that is typical in practical settings. 

\subsection{Contributions of the Paper}
\label{ssec:contributions}

The major contributions of this paper are as follows: 
\begin{itemize}
    \item We introduce a new algorithm for localization called Ordinal UNLOC. This algorithm combines rank aggregation, function learning, and unfolding optimization to solve the localization problem. 
    \item We show that using only one-bit data (ordinal, pairwise comparisons), we can accurately solve the localization problem. 
    \item We show through numerical simulations, that our method outperforms traditional localization approaches in rich scattering environments. 
    \item Through hardware experiments, we show that Ordinal UNLOC outperforms traditional localization. 
    \item We report that Ordinal UNLOC is only slightly worse off than fingerprinting-based methods, but with significant lower overhead.  \\ \\
\end{itemize} 

The rest of this paper is organized as follows. In Section \ref{sec:system_model}, the problem is described. The Ordinal UNLOC algorithm introduced in this paper to solve the localization problem is presented in Section \ref{sec:ordinal_UNLOC}. Simulation and hardware experiments are discussed in Section  \ref{sec:results_discussion}. Concluding remarks and future work are presented in Section \ref{sec:conclusions}.

\begin{figure}[t!bp]
	\centering
	\includegraphics[width=0.7\linewidth]{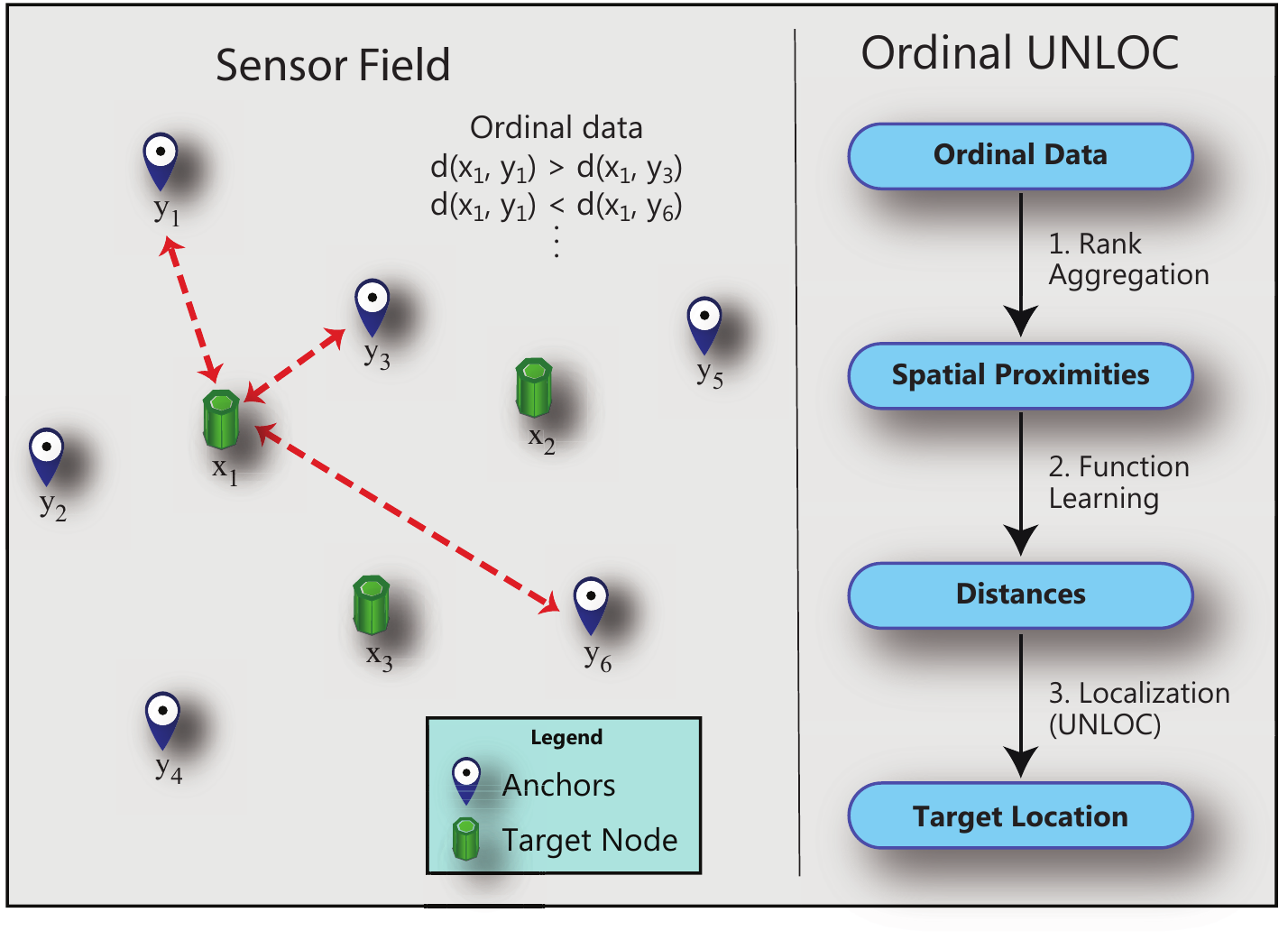}
	\caption{Sensor field (shaded in light gray) that contains anchors (sensors at known locations, denoted by blue and white pins) and target nodes (sensors at unknown locations, marked as green hexagonal prisms). Only comparative pairwise distances are known, giving rise to ordinal data. Ordinal UNLOC is used to estimate the location of target nodes based on ordinal data. The method contains three main steps: rank aggregation (Section~\ref{ssec:rank_aggregation}), function learning (Section~\ref{ssec:learningdistance}), and unfolding localization using approximate anchor-to-target distances (Section~\ref{ssec:UNLOC}).} 
	\label{Fig:sensor_field}
\end{figure}

\section{Related Work}
\label{sec:related_work}

Predominant research on indoor localization focuses on addressing the lack of infrastructure, by utilizing wireless sensor networks (WSNs). A given area is covered by placing a set of low-cost wireless sensors (called anchors). The anchors, together with the target node that is to be localized, form a sensor network via wireless communications (See Figure \ref{Fig:sensor_field}). Assuming that the distance between the anchors and the target can be reasonably estimated, such approximate anchor-to-target distances together with the known location of the anchors are used to estimate the location of the target. Under this framework, many methods have been developed for target localization based on noisy distance measures, including multilateration and triangulation \cite{PH03, Pa05, Wymeersch2009, WB12, ZB12, ZT13b, Zhang2016}, and linear and nonlinear optimization \cite{beck08, Chan09, Wang09, Moha11}. However, as  noted earlier, these methods are not suitable for practical indoor localization since the distances measured between the sensors, typically inferred from proxies such as time and power signals, are generally unreliable in indoor environments \cite{ozer16, chai16, altini10, kriz16, chintalapudi10, rida15, palumbo15, Tragas07, Gholami13, jin15, benedetto15, wang16, juri16, faragher15}.

In recent years, fingerprinting methods have become increasingly common for localization with superior accuracy \cite{faragher15, Zhuang2016, jin15, Dani2017, Khandker2020} where signal distributions are mapped to specific regions in space. Though there are performance advantages to using fingerprinting, it requires extensive mapping of the environment, \textit{a priori}. This presents a challenge in dynamic environments that change frequently. It is also necessary that the training database be sufficiently large to provide robustness to the algorithm. Several methods to mitigate these issues have been developed including neural networks \cite{altini10}, histogram interpolation to decrease the amount of sampling required \cite{Dani2017}, and sensor fusion \cite{chintalapudi10, faragher15, kriz16}. Methods to use the database for localization include Bayesian estimators \cite{faragher15}, and deep learning \cite{kriz16, altini10, Akino2020}.

In contrast to the work described above, we propose an algorithm that does not require detailed distance measurements, or rely on pretrained, precollected, and expensive methods such as with fingerprinting.  

\section{System Model}
\label{sec:system_model}
A standard setup of a sensor field for indoor localization is shown in Figure \ref{Fig:sensor_field}, where $N$ sensors are classified depending on whether their locations are known or unknown: $m$ sensors at known locations are called {\it anchors} and the $n$ sensors at unknown locations are {\it targets} (note that $N=n+m$).
We use $\bm{y}_i=[y_{1i},\dots,y_{qi}]^\top\in\mathbb{R}^{q}$ and  $\bm{x}_j=[x_{1j},\dots,x_{qj}]^\top\in\mathbb{R}^{q}$ to represent the coordinates of the $i$-th anchor ($i=1,\dots,m$) and those of the $j$-th target ($j=1,\dots,n$) in a $q$-dimensional space, respectively. Typically $q=2$ or $q=3$ but our analysis is valid for general $q$. We classify the distances between sensor pairs into three types: anchor-to-anchor, anchor-to-target, and target-to-target, which we represent using three matrices: $\mathbf{D}^{Y} = [d^{Y}_{ij}]_{m\times m}$, $\mathbf{D}^{X} = [d^{X}_{ij}]_{n\times n}$, and $\mathbf{D}^{YX} = [d^{YX}_{ij}]_{m\times n}$, respectively, where
$d^{Y}_{ij}=\|\bm{y}_i-\bm{y}_j\|,~d^{X}_{ij}=\|\bm{x}_i-\bm{x}_j\|,~d^{YX}_{ij}=\|\bm{y}_i-\bm{x}_j\|.$
We also define an aggregated distance matrix $\mathbf{D}=[d_{ij}]_{N\times N}$ as
\begin{equation}
\mathbf{D} =
\begin{pmatrix}
  \mathbf{D}^Y & \mathbf{D}^{YX} \\
  \mathbf{D}^{XY} & \mathbf{D}^X
\end{pmatrix},
\end{equation}
where $\mathbf{D}^{XY}$ is the transpose of $\mathbf{D}^{YX}$. When we refer to the sensors without specifying the type, the first $m$ sensors are anchors and the rest ($n$) are targets, so that sensor $i$ for $1\leq i\leq m$ refers to the $i$-th anchor, whereas sensor $m+j$ refers to the $j$-th target node ($j=1,\dots,n$). 

Since the anchor locations are known, the matrix $\mathbf{D}^{Y}$ can be directly obtained. However, for many practical settings such as in indoor environments, the rest of the block matrices $\mathbf{D}^X$, $\mathbf{D}^{YX}$, and $\mathbf{D}^{XY}$ are generally unavailable~\cite{WB12, ZB12, ZT13b, banavar2018}. Here, we take a different approach. Instead of using unreliably measured distances, we consider comparisons of distances (or their proximity measures) which give rise to ordinal comparison data, which we describe as follows.
For each sensor triplet $(i,j,k)$, we compare the distance measurements or proxies from sensor $i$ and from sensor $j$ to the reference sensor $k$, respectively, to determine which sensor (between $i$ and $j$) is closer to $k$. We denote the binary outcome of such a pairwise comparison as $z^{(k)}_{ij}$, which generally depends on some unknown function $f$ of the actual distances $d_{ik}$ and $d_{jk}$, represented as
\begin{equation}\label{eq:generalcomp}
z^{(k)}_{ij} = f\left(d_{ik},d_{jk},\xi^{(k)}_{ij}\right),
\end{equation}
where $\xi^{(k)}_{ij}$ denotes noise in such a comparison. Due to skew symmetry constraints ($i$ being closer to $k$ than $j$ is the same as $j$ being farther away from $k$ than $i$, and vice versa), we assume that $\xi^{(k)}_{ij}=-\xi^{(k)}_{ji}$ and require that $f(d,d',\xi)=-f(d',d,-\xi)$, which ensures that $z^{(k)}_{ij}=-z^{(k)}_{ji}$ ($\forall i,j,k$).
For the majority of the paper, we focus on {\it ordinal} comparisons, defined by a ``thresholding" function
\begin{equation}\label{eq:ordinalcomp}
f(d,d',\xi)=\sgn (d-d'+\xi)\in\{-1,1\},
\end{equation}
where $\sgn(\cdot)$ is the signum function \cite{spanier1987atlas}.
Note that this is a special case of the logistic function $f(d,d',\xi)=1/\left[1+e^{-\beta (d-d'+\xi)}\right]$ in the limit of the parameter $\beta\rightarrow\infty$.
Furthermore, in practice one might only have measurements of physical quantities (``signals") that serve as proximity of distance, in which case 
\begin{equation}\label{eq:ordinalcomp2}
f(d,d',\xi)=\sgn (s(d)-s(d')+\xi)\in\{-1,1\},
\end{equation}
where $s$ is a (monotonic) function of distance, such as power loss or time of flight of signals between sensors.

\begin{figure}[tbp]
	\centering
	\includegraphics[width=0.5\linewidth]{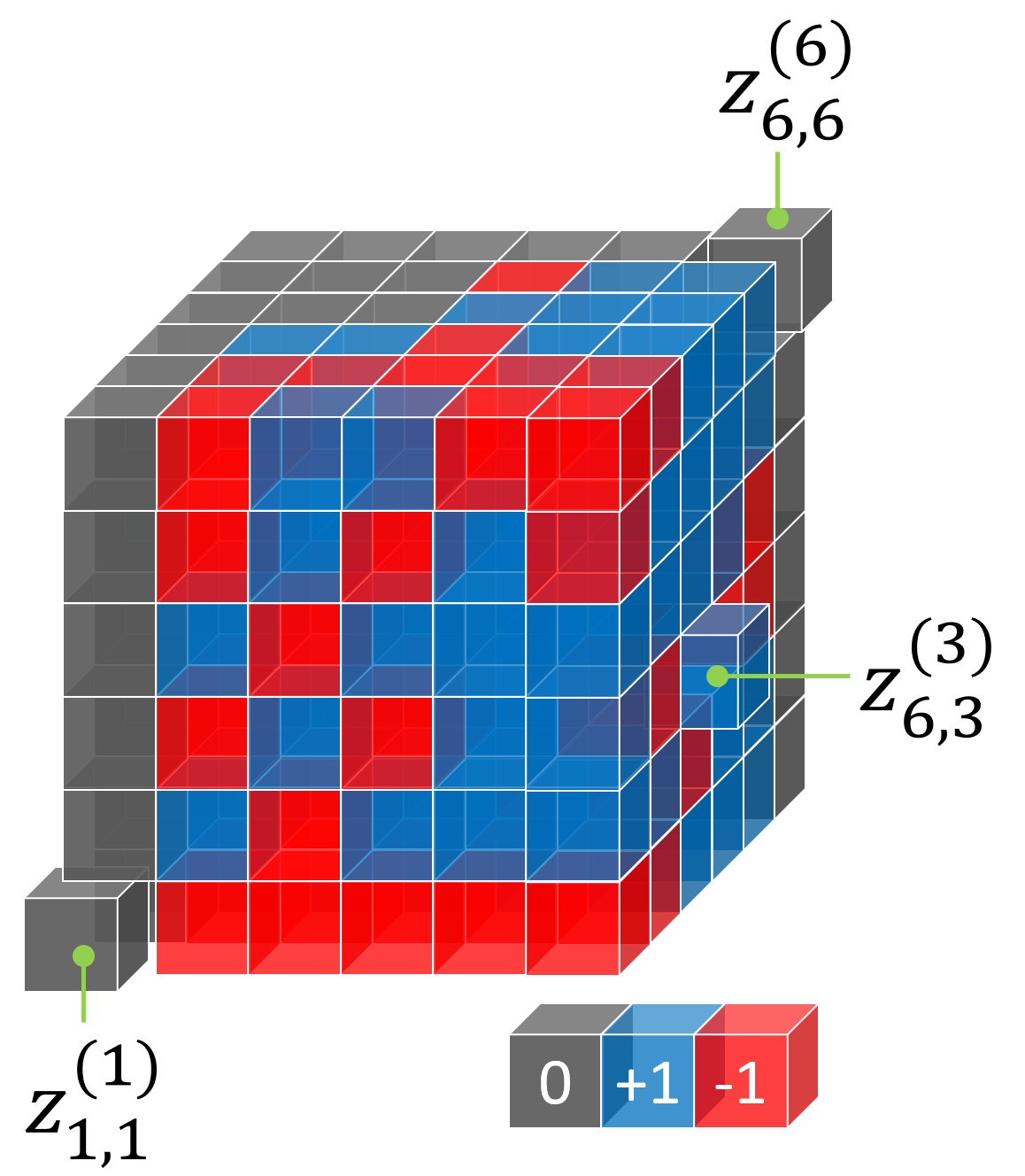}
	\caption{Visual illustration of the $\mathbb{Z}$ tensor as introduced in Section \ref{sec:system_model}. Each small cube in the larger cube represents one entry in the tensor, with a +1 represented by blue, -1 by red, and zero by gray. This particular tensor was generated during one instance of the Monte-Carlo simulation that resulted in Figure \ref{fig:mse_v_anchors}. }
	\label{fig:tensor}
\end{figure}

The collection of all pairwise comparisons form $N$ square-shaped matrices: $\{\mathbf{Z}^{(k)}\}$, where $\mathbf{Z}^{(k)}=[z^{(k)}_{ij}]_{N\times N}$ ($k=1,\dots,N$), which we refer to as pairwise comparison matrices, can be combined into single tensor $\mathbf{Z}$ (see Figure \ref{fig:tensor} for an example). 

\section{Ordinal UNLOC}
\label{sec:ordinal_UNLOC}

Our localization algorithm, which we term {\it \underline{Ordinal} \underline{Un}folding-based \underline{Loc}alization (Ordinal UNLOC)}, consists of three main steps (see Fig. \ref{Fig:sensor_field}, right panel). First, from the ordinal comparison data, we apply {\it rank aggregation}  to obtain a set of ``dissimilarities", which, for a given reference sensor, $k$, assigns a score to each sensor that serves as a proxy to its distance to the $k$-th sensor according to the ordinal data and rankings. Such sets of spatial proximities (scores) are not unique, since any shifting and monotonic scaling preserves the ranking of the scores. Consequently, the scores obtained in this step cannot be directly used as approximate anchor-to-target distances for localization.  

Next, the estimated distance proximities among the anchors, together with the known anchor-to-anchor distances, are used to fit functions that best transform proximities into distances, and such functions are then applied to obtain an estimation of anchor-to-target distances.

Finally, given the locations of anchors and estimated anchor-to-target distances, we formulate a multidimensional unfolding optimization problem, the solution of which provides an estimate of the location of the target.

In what follows, we describe each of these steps. 

\subsection{Rank aggregation from ordinal data}
\label{ssec:rank_aggregation}

The first step in Ordinal UNLOC is to use a rank aggregation method to infer spatial proximities from the given ordinal data. For each reference sensor $k$, we seek a set of scores, denoted by $\bm{\psi}^{(k)}=[\psi_{1k},\dots,\psi_{Nk}]^\top$, that serve as proximities and ideally preserves the ordinal data as constraints, such that $\psi_{ik}>\psi_{jk}$ if and only if $d_{ik}>d_{jk}$ (or in other words, $z^{(k)}_{ij}=1$). Note that this is equivalent to requiring the ranking of the entries in each $\bm{\psi}^{(k)}$ be identical (or as close as possible) to the ranking of the entries of the $k$-th column of $\mathbf{D}$, which we denote as $\bm{d}^{(k)}=[d_{1k},\dots,d_{Nk}]^\top.$

The problem of inferring the ranking or scoring of a set of items from their pairwise comparisons is commonly known as ``rank aggregation'' \cite{chen2013pairwise,ammar2012efficient}, interpreting each pairwise comparison as assigning a local ranking between two items, with the goal of obtaining an aggregated global ranking that preserves these local rankings as much as possible.
Several methods are available for solving a rank aggregation problem, most of which compute a {\it score} for each item based on the collection of ordinal data \cite{negahban2016rank}. 

Here, we adopt a common, easy-to-implement, and effective method, referred to as HodgeRank \cite{jiang2011statistical} or simply least squares (LS) ranking. The idea is to formulate a linear least squares problem using the ordinal data as inputs, the solution of which gives the estimated spatial proximities and approximately preserves distance orderings from the comparison data. Consider an arbitrary enumeration of the set of all ordered pairs of sensors, where the $\ell$-pair is denoted by $(i_\ell,j_\ell)$, defining a set
\begin{eqnarray}\label{eq:edgeset}
E&=&\{(i_\ell,j_\ell):\ell=1,\dots,M\}\nonumber\\
&=&\{(i,j)\in\mathbb{N}^2:1\leq i<j\leq N\},
\end{eqnarray}
where $M=N(N-1)/2$. Thus, $E$ can be interpreted as the edge set of a complete graph of $N$ nodes, and $\{(i_\ell,j_\ell)\}$ is an ordered list of all the edges. From this, we define the corresponding {\it incidence matrix}, $\mathbf{B} = [b_{\ell,q}]_{M\times N}$ \cite{bondy1976graph}, where
\begin{equation}\label{eq:incidencematrix}
b_{\ell,q}=
\begin{cases}
1,~\mbox{if $q=i_\ell$},\\
-1,~\mbox{if $q=j_\ell$},\\
0,~\mbox{otherwise}.
\end{cases}
\end{equation}
For each sensor $k$, the ordinal data in matrix $\mathbf{Z}^{(k)}$ can be effectively represented by a column vector
\begin{equation}\label{eq:zk}
\bm{z}^{(k)}=[z^{(k)}_{i_1,j_1},\dots,z^{(k)}_{i_M,j_M}]^\top\in\mathbb{R}^{M},
\end{equation}
following the same enumeration of pairs as in the incidence matrix. Then, the LS ranking method solves a linear least squares problem to yield
\begin{equation}\label{eq:LSranking}
\bm{\psi}^{(k)} = \argmin_{\bm\psi\in\mathbb{R}^N,\bm{1}^\top\bm\psi=0} \| \mathbf{B} \bm{\psi} - \bm{z}^{(k)} \|.
\end{equation}
Such a solution can be computed in several ways, including for example, using normal equations or singular value decomposition \cite{golub2012matrix}, to produce
\begin{equation}
\bm{\psi}^{(k)} = (\mathbf{B}^\top \mathbf{B})^\dagger \mathbf{B}^\top\bm{z}^{(k)} + c\bm{1},
\end{equation}
where $(\cdot)^\dagger$ denotes the pseudoinverse of a matrix \cite{golub2012matrix}, the matrix $\mathbf{B}^\top \mathbf{B}$ is also known as the graph Laplacian and is generally noninvertible, and the constant $c$ is chosen such that the condition $\bm{1}^\top\bm{\psi}^{(k)} = 0$ is satisfied.
Finally, we collect the vectors $\bm\psi^{(k)}$ into a matrix of spatial proximities
\begin{equation}
\bm\Psi = [\bm\psi^{(1)},\dots,\bm\psi^{(N)}].
\end{equation}
We partition this matrix in the same way as the aggregated distance matrix $\mathbf{D}$, so that
\begin{equation}
\bm\Psi_{N\times N}=
\begin{pmatrix}
\bm\Psi^Y_{m\times m} & \bm\Psi^{YX}_{m\times n} \\
\bm\Psi^{XY}_{n\times m} & \bm\Psi^X_{n\times n}
\end{pmatrix},
\label{eqn:psi_matrix}
\end{equation}
where $\bm\Psi^{XY}$ is the transpose of $\bm\Psi^{YX}$.

\subsection{Function Learning: estimating distances from spatial proximities}
\label{ssec:learningdistance}

The second step of Ordinal UNLOC is to estimate the unknown anchor-to-target distances from the matrix of spatial proximities, $\bm\Psi$, obtained by rank aggregation in (\ref{eqn:psi_matrix}), together with the known anchor-to-anchor distances. In particular, for each sensor $k$ that is an anchor (so $1\leq k\leq m$), we learn a function $g_k$ that maps the estimated proximities with respect to anchor $i$ into known distances $d_{ik}$, so that $d_{ik}\approx g_k(\bm\psi^{(k)}_i)$ ($i=1,\dots,m$). Since $d_{ik}$ is only known for $1\leq i\leq m$ (i.e., distances between pairs of anchors), the function $g_k$ needs to be inferred from the $k$-th column of the anchor-to-anchor distance matrix $\mathbf{D}^Y$ and that of the dissimilarity matrix $\bm\Psi^Y$, which we denote as $\bm{d}^{Y}_k$ and $\bm\psi^{Y}_k$, respectively.

We express $g_k$ using a basis expansion. Using the standard polynomial basis, we have the representation
\begin{equation}
g_k(\psi)=\sum_{l=0}^{\infty}c^{(k)}_{l}\psi^{l}.
\label{eqn:poly_basis}
\end{equation}
To preserve the ordering among the dissimilarities, we additionally require that $g_k$ be a monotonic function. Under this additional constraint, we here consider a truncated series to the first order so that $g_k$ becomes just a linear function: 
\begin{equation}\label{eq:glinear}
g_k(\psi)=c^{(k)}_0+c^{(k)}_1\psi,
\end{equation}
where the coefficients $c^{(k)}_0$ and $c^{(k)}_1$ are to be determined from the vectors $\bm{d}^{Y}_k$ and $\bm\psi^{Y}_k$. This can be done via solving a linear regression problem, using standard least squares, for example, producing
\begin{equation}
\bm{c}^{(k)}=[c^{(k)}_0,c^{(k)}_1]^\top = \argmin_{\bm{c}\in\mathbb{R}^2,c_1>0}\|[\bm{1},\bm\psi^{Y}_k]\bm{c}-\bm{d}^{Y}_k\|,
\end{equation}
for $k=1,\dots,m$ (every anchor).

From these coefficients, we compute an estimate of the distance from the $k$-th anchor to the targets, using the formula
\begin{equation}
\tilde{\bm{d}}^{XY}_k = [\tilde{d}^{XY}_{1k},\dots,\tilde{d}^{XY}_{nk}]^\top = c^{(k)}_0 + c^{(k)}_1\bm\psi^{XY}_k.
\end{equation}
Repeating the procedure for all the anchors, $k=1,\dots,m$, produces a preliminary estimate of the anchor-to-target distance matrix, $\tilde{\mathbf{D}}^{XY}$ (or equivalently, $\tilde{\mathbf{D}}^{YX}$, by transposing $\tilde{\mathbf{D}}^{XY}$).

Before moving on to the next step of localization, it is important to recalibrate the estimated anchor-to-target distances, for the following reason. Consider the $j$-th target node ($j\in\{1,\dots,n\}$), whose distances to the anchors are preliminarily estimated to form the $j$-th column of $\tilde{\mathbf{D}}^{YX}$, denoted by
$\tilde{\bm{d}}^{YX}_j=[\tilde{d}^{YX}_{1j},\dots,\tilde{d}^{YX}_{mj}].$
For each anchor $k\in\{1,\dots,m\}$, the entry $\tilde{d}^{YX}_{kj}$ is obtained from its corresponding function $g_k$, which generally differs from one anchor to another. The error in fitting $g_k$ can cause the ordering of $\tilde{d}^{YX}_{kj}$'s to differ from the ordering of the true distances $d^{YX}_{kj}$, and even different from the estimated proximities $\psi^{YX}_{kj}$. Therefore, for each target node $j$, to best ensure that the estimated anchor-to-target distances preserve the ordering of estimated proximities, we re-fit a linear function $g_{m+j}$ in the same form as (\ref{eq:glinear}), but now with the goal of constructing a monotonic function that transforms the anchor-to-target proximities to the preliminary estimate of anchor-to-target distances, in place of the unknown distances, giving:
\begin{align}
\bm{c}^{(m+j)} &= [c^{(m+j)}_0,c^{(m+j)}_1]^\top \nonumber \\
&= \argmin_{\bm{c}\in\mathbb{R}^2,c_1>0} \|[\bm{1},\bm\psi^{YX}_{j}]\bm{c}-\tilde{\bm{d}}^{YX}_{j}\|,
\end{align}
for $j=1,\dots,n$, where $\bm{\psi}^{YX}_j=[\psi^{YX}_{1j},\dots,\psi^{YX}_{mj}]$. Finally, we use $g_{m+j}$ to recalibrate the estimated anchor-to-target distances, as
\begin{equation}\label{eq:newdist}
\hat{\bm{d}}^{YX}_j = c^{(m+j)}_0 + c^{(m+j)}_1\bm\psi^{YX}_j,
\end{equation}
for $j=1, \dots, n$, thus forming the estimated anchor-to-target distance matrix $\hat{\mathbf{D}}^{YX}$ and its transpose $\hat{\mathbf{D}}^{XY}$.

\subsection{Unfolding localization from distance measures}
\label{ssec:UNLOC}

For each target node $j$ (i.e., sensor $m+j$, where $1\leq j\leq n$), the estimated anchor-to-target distances $\hat{\mathbf{D}}^{YX}$ together with the locations of the anchors $\{\bm{y}_{i}\}$ can be used to infer the location of the target $\bm{x}_j$. We achieve this by formulating an unfolding optimization, with cost function \cite{Sun2018}
\begin{equation}
J(\bm{x},\mathbf{Y}; \bm{\delta}) = \sum_{i=1}^{m} \left(\|\bm{x} - \bm{y}_i\|^2 - \delta_{i}\right)^2,
\end{equation}
where the matrix $\mathbf{Y}=[\bm{y}_1,\dots,\bm{y}_m]$, $\bm\delta = [\delta_1,\dots,\delta_m]^\top$, and the solution is termed UNLOC \cite{Sun2018}.
Applying UNLOC to each column of the estimated distance matrix $\hat{\mathbf{D}}^{YX}$ leads to the estimated location of all the targets. That is, we compute, for each $j=1,\dots,n$,
\begin{equation}
\hat{\bm{x}}_j = \argmin_{\bm{x}} J(\bm{x}, \mathbf{Y}; \hat{\bm{d}}^{YX}_j). 
\label{eq:UNLOC_optimization}
\end{equation}
This can be solved in several ways, including those discussed in \cite{beck08}, a global optimization routine \cite{cox2000multidimensional}, or generalized to include weights \cite{Sun2018}. 

\subsection{Computational Complexity}
\label{ssec:complexity}

In this section, we discuss the computational complexity of Ordinal UNLOC with reference to traditional localization methods. In terms of computational complexity, the final step of Ordinal UNLOC is a traditional multilateral algorithm such as the one discussed in \cite{Sun2018}. Ordinal UNLOC is, therefore, more complex than traditional ToA/RSS-based systems. However, while the traditional methods require accurate distance or time estimates, our approach does not. The cost in terms of computational complexity is therefore mitigated by the fact that we can obtain accurate localization estimates even in rich scattering environments.

\section{Experiments and Results}
\label{sec:results_discussion}

To validate Ordinal UNLOC and verify its effectiveness, we conduct a variety of numerical simulations. In each such simulation, we place $m$ anchors as well as a target uniformly at random in $[0,1]^2$ and obtain ordinal comparison data from the threshold model \eqref{eq:generalcomp}-\eqref{eq:ordinalcomp} using Gaussian noise $\mathcal{N}(0,\sigma^2)$. Specifically, for $m=5$, we show an example of the $\mathbf{Z}$ tensor in Fig. \ref{fig:tensor}, for one realization of noise.

\begin{figure}[tbp]
	\centering
	\includegraphics[width=0.4\linewidth]{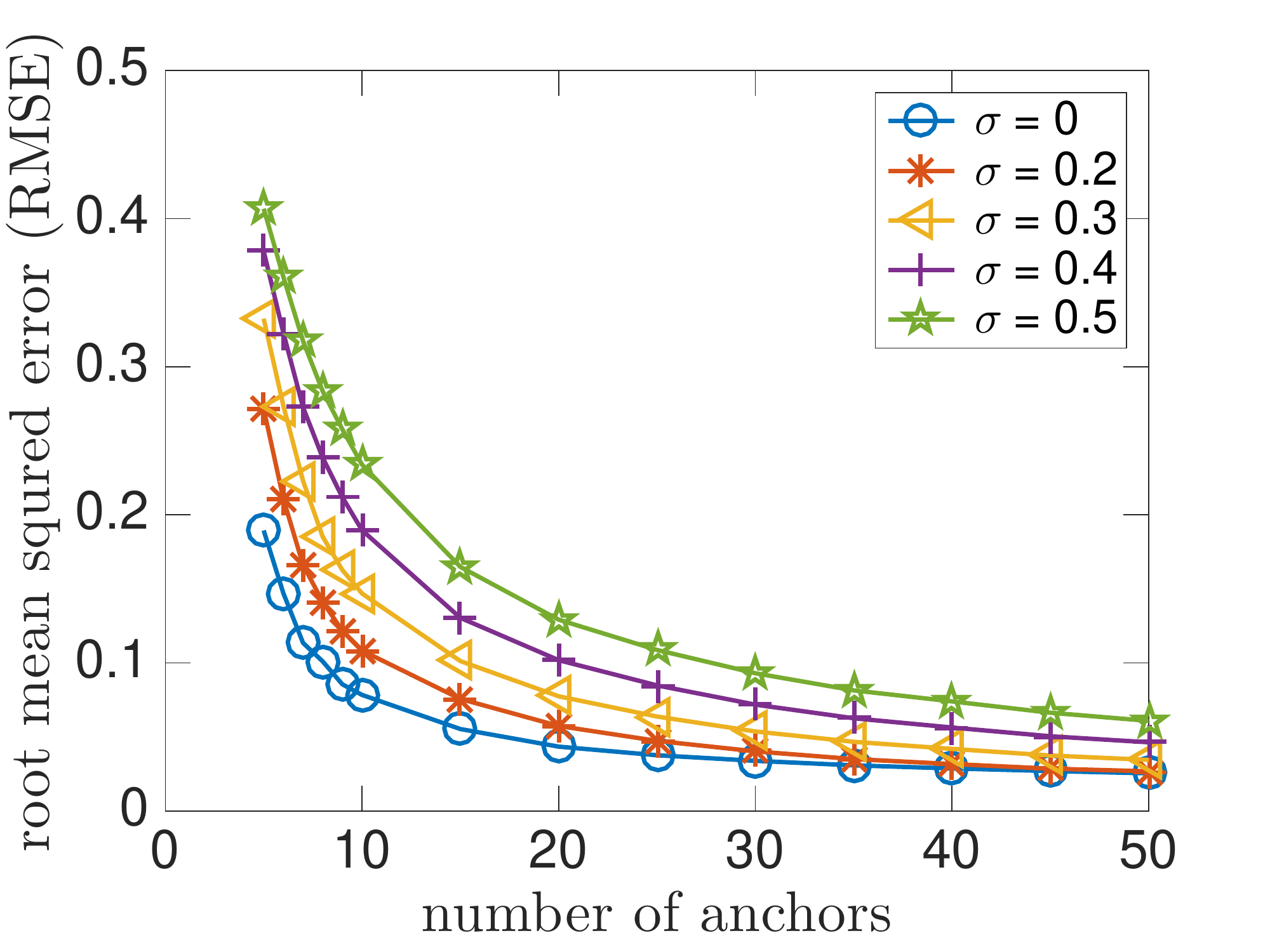}
	\caption{Dependence of localization error on the number of anchors. We plot the root mean squared error (RMSE) of Ordinal UNLOC as a function of the number of anchors $m$ and data collected at noise levels from $\sigma = 0$ to $\sigma = 0.5$. The results are averaged over 5000 simulations and the locations of the sensors are chosen randomly for every simulation run.}
	\label{fig:mse_v_anchors}
\end{figure}

\subsection{Benchmark tests: effects of the number of anchors and noise}
\label{ssec:benchmark_tests}

As the result of the Monte-Carlo simulations, as shown in Fig. \ref{fig:mse_v_anchors}, we found that the root-mean-squared-error (RMSE) of localization decreases rapidly as the number of anchors increases. Because ordinal comparisons inevitably introduce information loss about the exact target location, a non-diminishing localization error is expected, and is indeed observed in our simulations. Both the quick improvement with increasing anchors and error saturation is found across a range of noise levels, suggesting robustness of Ordinal UNLOC. 

\begin{figure}[tbp]
	\centering
	\includegraphics[width=0.4\linewidth]{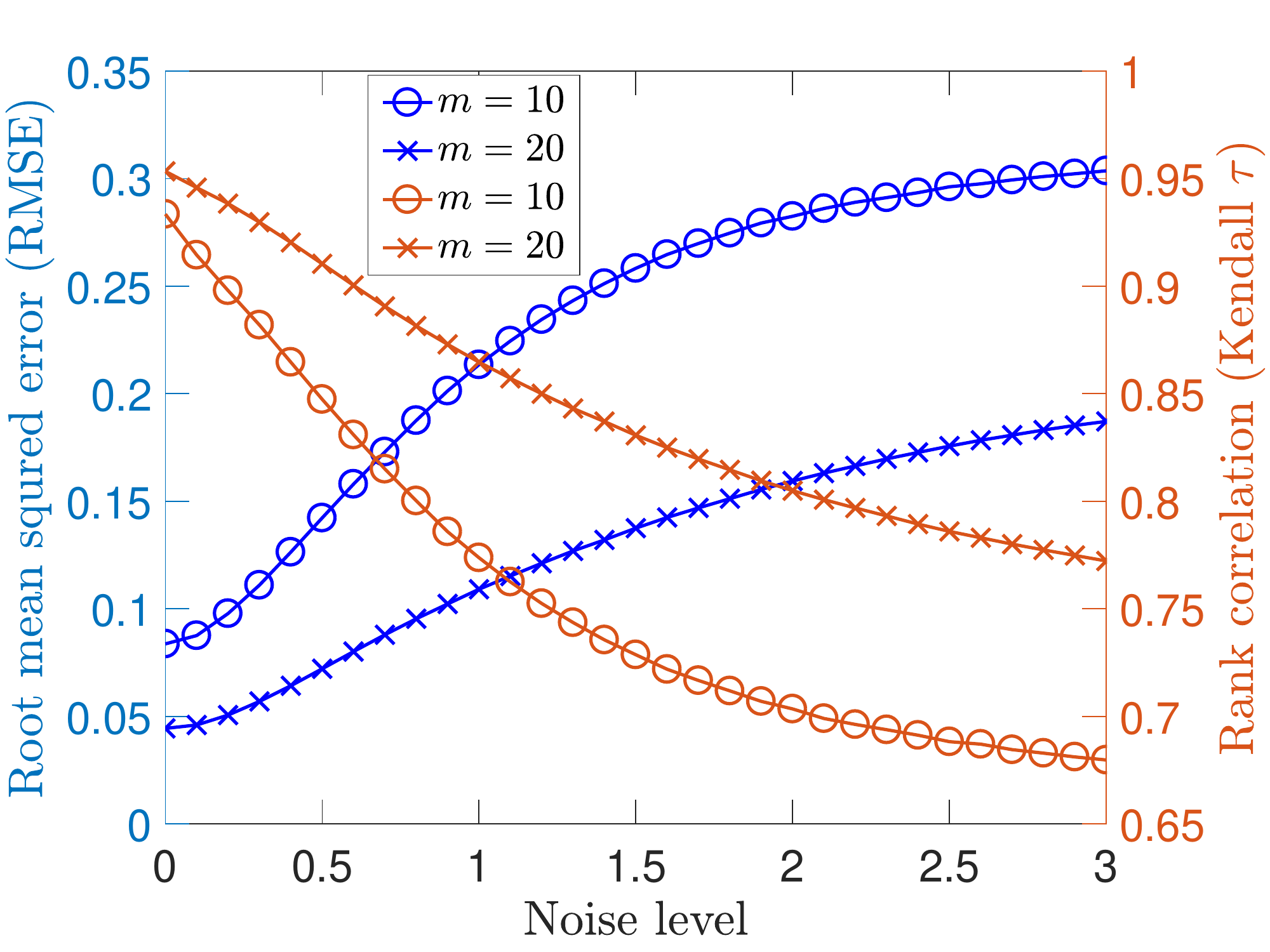}
	\caption{Dependence of localization error and rank correlation (Kendall $\tau$) on the level of noise. Here we fix the number of anchors $m$ (either $m=10$ or $m=20$), and use a single target node ($n=1$). We plot the RMSE of localization and the rank correlation of $\mathbf{D}$ and $\hat{\mathbf{D}}$ as a function of the noise standard deviation $\sigma$.} 
	\label{fig:mse_v_rank_correlation}
\end{figure}

To further understand the non-perfectness of localization with ordinal constraints, we compare the RMSE of localization with the rank correlation between $\mathbf{D}$ and $\hat{\mathbf{D}}$, by plotting both versus increasing noise levels. The results, shown in Fig. \ref{fig:mse_v_rank_correlation}, validate our hypothesis that increasing the number of anchors improves performance, and increasing noise deteriorates performance. It can also be seen that as the rank correlation between the distance matrices decreases, that is, as the estimates of the cross-distances between the anchors and the targets reduce its quality, the error in localization increases. 

\subsection{Localization under physical transmission models}
\label{ssec:communications_models}

Although we have shown the effectiveness of Ordinal UNLOC in simulations using ordinal data, such ordinal data come directly from comparing distances under white noise. However, effective  noise in practice is expected to deviate from this idealized scenario. For example, a sensor typically transmits and receives signals that are distance-dependent. For a method to be potentially useful, it must be able to deal with more realistic physical transmission models, which is the focus of our next set of numerical experiments. We consider two common models: received signal strength and time of arrival. 

\subsubsection{Received Signal Strength (RSS)}
\label{sssec:RSS}

Given a signal transmitted with power, $P_{T}$, the received power, $P_R$, is given by \cite{banavar2018} 
\begin{equation}
    P_R = P_{T} \alpha d^{-G},
    \label{eqn:receive_power}
\end{equation}
where $\alpha$ is a constant that depends on the hardware and signal characteristics and $G$ is the path-loss exponent, which equals two in free space and is typically larger in indoor environments \cite{Ba2005}. Given a sensor field, we now assume that signals are transmitted with strength $P_{T}$ and the received signal strength (RSS) is given according to (\ref{eqn:receive_power}). To mimic realistic challenges of unknown environments, we conduct a simulation where $G$ is drawn randomly in the interval $[a,b]$ with $a=2$ and $b=6$ \cite{juri16}.

\begin{figure}[tbp]
	\centering
	\includegraphics[width=0.4\linewidth]{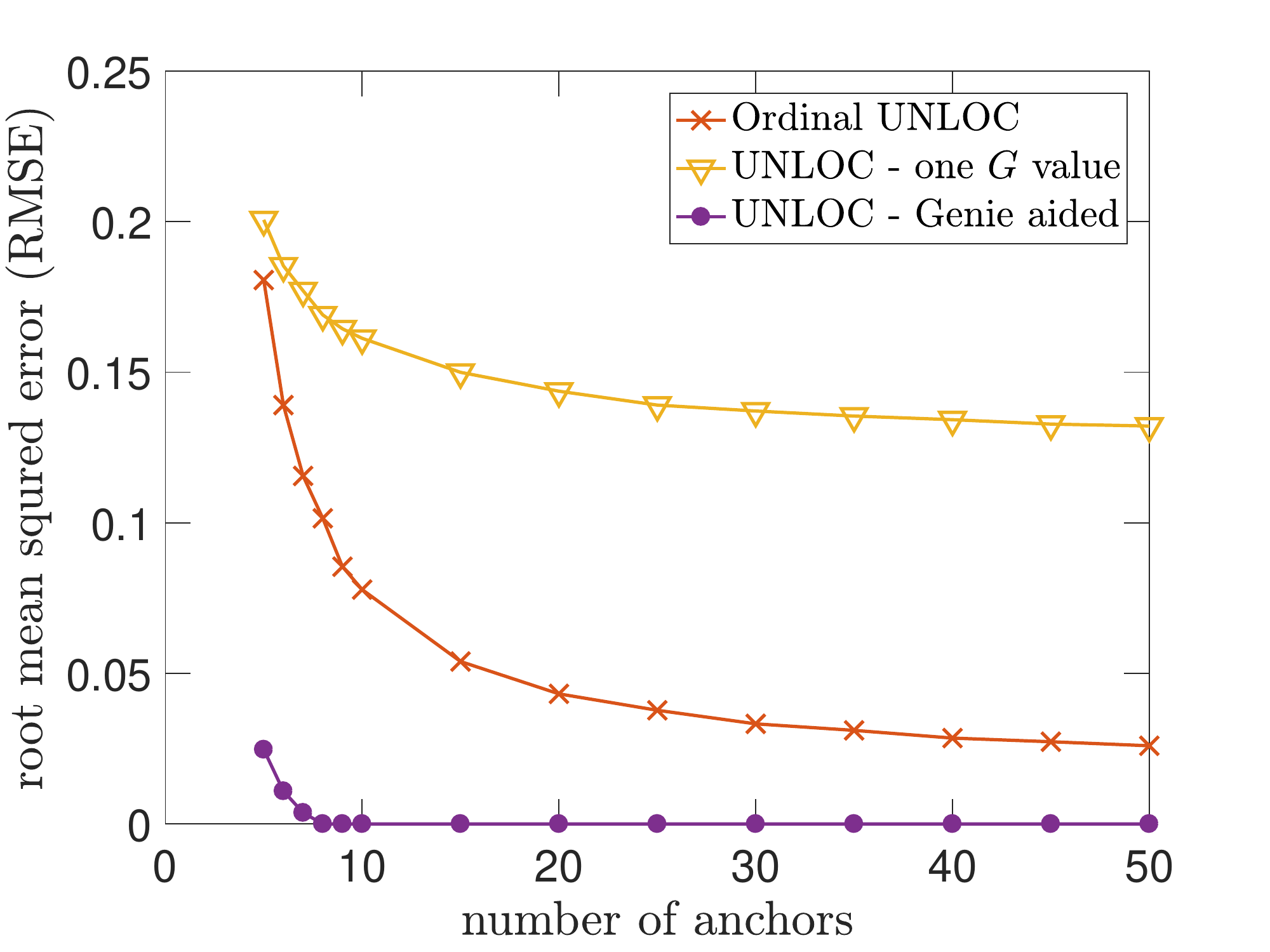}
	\caption{We compare UNLOC \cite{Sun2018} with Ordinal UNLOC. Each transmit receive link sees a different value of $G$ (case (iii)). For Ordinal UNLOC, we directly compare the received power values. We then estimate the distances between anchors and the target using (\ref{eqn:receive_power}). In one case, a fixed $G$, that is derived using calibration is used. This is compared against the ideal ``Genie-aided'' case, where instantaneous $G$ values of each link are available. While the Genie-aided localization performs the best, Ordinal UNLOC directly with received power outperforms UNLOC in a realistic setting.}
	\label{fig:mse_RSS}
\end{figure}

We perform three numerical experiments. First, we run Ordinal UNLOC on the RSSI values directly. This is compared with two other scenarios, where the distance is estimated from the RSSI values first using (\ref{eqn:receive_power}), and then, UNLOC \cite{Sun2018} is applied in two ways. In the first scenario, a single estimate of the path loss exponent, $G$, is available for the entire environment, as is typically the case \cite{juri16, banavar2018}. In the second case, we assume Genie-aided calibration using (\ref{eqn:receive_power}), where the path loss exponent for each link available at all times. Performance for these three cases, where we plot localization error (MSE) versus the number of anchors, is shown in Fig. \ref{fig:mse_RSS}. Clearly, having all information in the Genie-aided  case provides the best performance. In the more realistic cases, Ordinal UNLOC outperforms traditional localization even though some information about the path loss exponent is known to UNLOC, but no information is available to Ordinal UNLOC except for the raw RSS values. 

\subsubsection{Time of Arrival (TOA)}
\label{sssec:TOA}

\begin{figure}[tbp]
    \centering
	\includegraphics[width=0.4\linewidth]{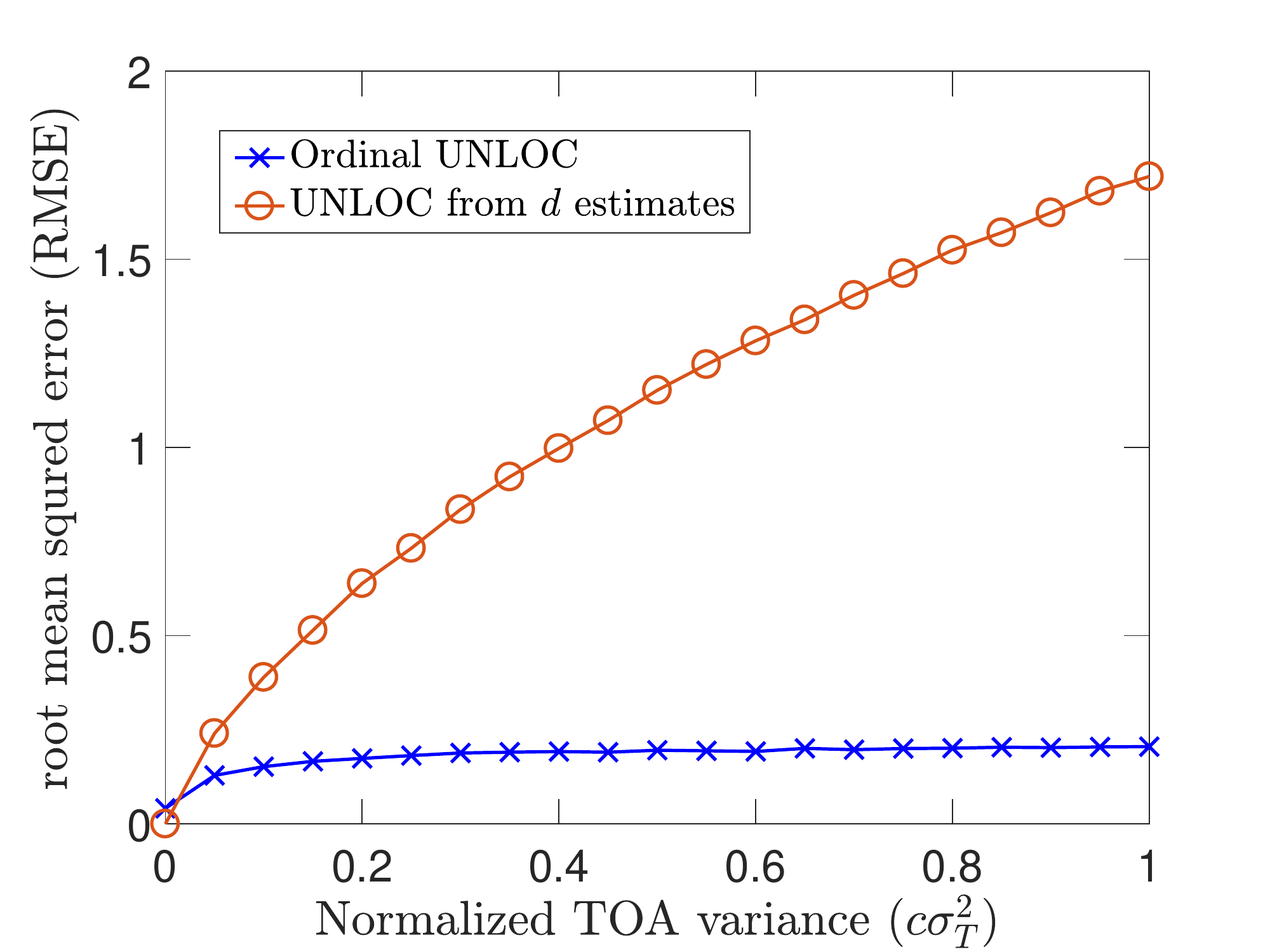}
	\caption{Comparing Ordinal UNLOC and UNLOC for TOA-based localization; $m = 20$. For Ordinal UNLOC, the time-of-flight values are directly compared to construct $\mathbf{Z}$. For UNLOC, first, the distance estimates between the anchors and the targets are estimated from time-of-flight values using \ref{eqn:TOA_model}, and the location is then estimated. As the uncertainty in time-of-flight values increases, the UNLOC performance deteriorates much faster than Ordinal UNLOC.}
	\label{fig:mse_TOA}
\end{figure}

In order to perform ranging using time of arrival, the signal propagation time between the transmitter and the receiver needs to be measured. In indoor environments, the time estimate is noisy, and may be modeled using a random variable as
\begin{equation}
    \hat{\tau} \sim \mathcal{N} \left( \frac{d}{c}, \sigma_{T}^{2} \right),
    \label{eqn:TOA_model}
\end{equation}
where $d$ is the true distance between the transmitter and the receiver, $c$ is the speed of propagation, and $\sigma_{T}^{2}$ is the variance of the TOA estimates on the channel \cite{Zhang2016, ZT13a, ZT13b}. We perform ordinal UNLOC on the TOA measurements, where the time estimates for each transmit-receive pair are drawn from the TOA model described in (\ref{eqn:TOA_model}). We enlarge the sensor field to a square of size 200 $\times$ 200 and use the normalized variance in order to better show the effects of variations in the TOA measurements \cite{Zhang2016, Zhang2018}. We plot the localization error vs the normalized variance $c \sigma_{T}^{2}$ for different number of anchors in Figure \ref{fig:mse_TOA}. 

\subsection{Hardware Experiments}
\label{ssec:hardware_expts}

Experiments were designed to validate our algorithm. Android devices were used as transceivers and placed in a lab environment. Four Android devices were designated as anchors and placed in fixed positions. One Android device was designated as the target and placed at a location unknown to the system. Bluetooth links were established between all the devices and received signal strength indication (RSSI) data was collected. The Ordinal UNLOC algorithm was applied to the collected data.

\begin{figure}[htbp]
	\centering
	\includegraphics[width=0.5\linewidth]{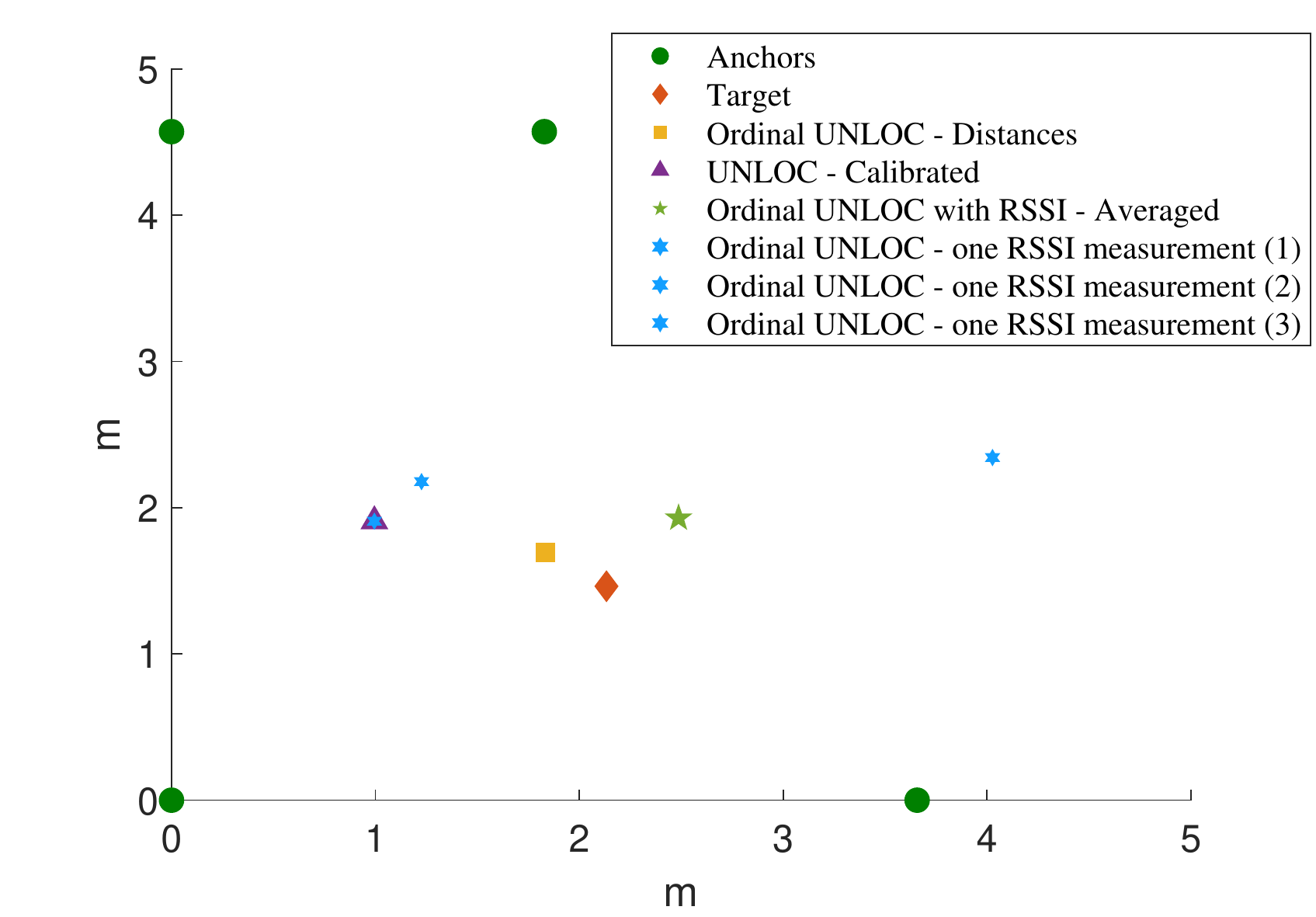}
	\caption{Hardware experiment to validate Ordinal UNLOC. Four anchors are placed  at  the  approximate  corners  of  a  rectangle  and  of many experiments conducted, two  targets  placed within  the  anchor  space are shown here.  Here, one of the targets is localized; the other target is individually localized in Figure \ref{Fig:config2_m} and both targets are simultaneously localized in Figure \ref{Fig:config3_m}. Ordinal  UNLOC  is  applied  on  both the  actual  cross-distances  and  the RSSI  values.  UNLOC  is  applied  on    distance  values derived from RSSI values. Ordinal UNLOC with no calibration outperforms UNLOC  with  calibrated  distance  estimates.  Averaging  all  Ordinal  UNLOC estimates with individual RSSI values provides an accurate location estimate, validating the Ordinal UNLOC algorithm.
	} 
	\label{Fig:config1_m}
\end{figure}

For the hardware experiment,  anchors were placed approximately in the corners of a rectangle of sides 4m $\times$ 5m. To avoid symmetry, that may play role in the results, the anchors were not placed exactly in the corners, but were moved away slightly. Within the convex hull of the four anchors, targets were placed in 20 separate locations for evaluation. Since the data was collected in a lab environment, due to multipath fading \cite{Goldsmith2005}, some paths are more reflective than others. Therefore, for each target location, from the over 10,000 measurements we collected, we reduced the number of localization attempts to 100, by only selecting the paths with large line-of-sight power using the Ricean $K$ factor estimator described in \cite{TA2003}. Once this was accomplished, the reduced set of RSSI values was used to estimate the location of each target 100 times, by using these RSSI values directly with Ordinal UNLOC. 

\begin{figure}[htbp]
	\centering
	\includegraphics[width=0.5\linewidth]{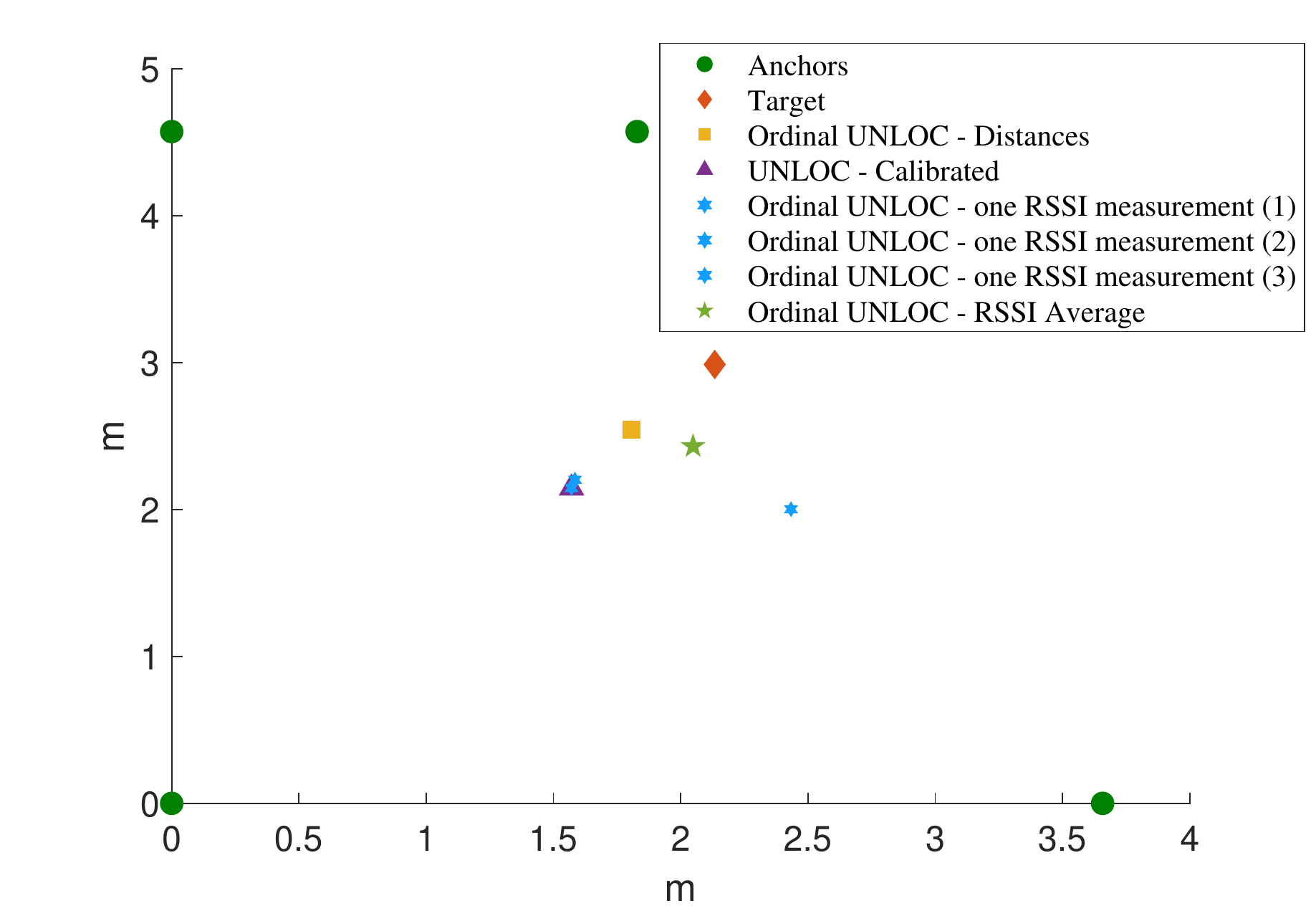}
	\caption{Hardware experiment to validate Ordinal UNLOC. Four anchors are placed  at  the  approximate  corners  of  a  rectangle  and  of many experiments conducted, two  targets  placed within  the  anchor  space are shown here.  Here, one of the targets is localized; the other target is individually localized in Figure \ref{Fig:config1_m} and both targets are simultaneously localized in Figure \ref{Fig:config3_m}. Ordinal  UNLOC  is  applied  on  both the  actual  cross-distances  and  the RSSI  values.  UNLOC  is  applied  on    distance  values derived from RSSI values. Ordinal UNLOC with no calibration outperforms UNLOC  with  calibrated  distance  estimates.  Averaging  all  Ordinal  UNLOC estimates with individual RSSI values provides an accurate location estimate, validating the Ordinal UNLOC algorithm.
	} 
	\label{Fig:config2_m}
\end{figure}

\begin{figure}[htbp]
	\centering
	\includegraphics[width=0.5\linewidth]{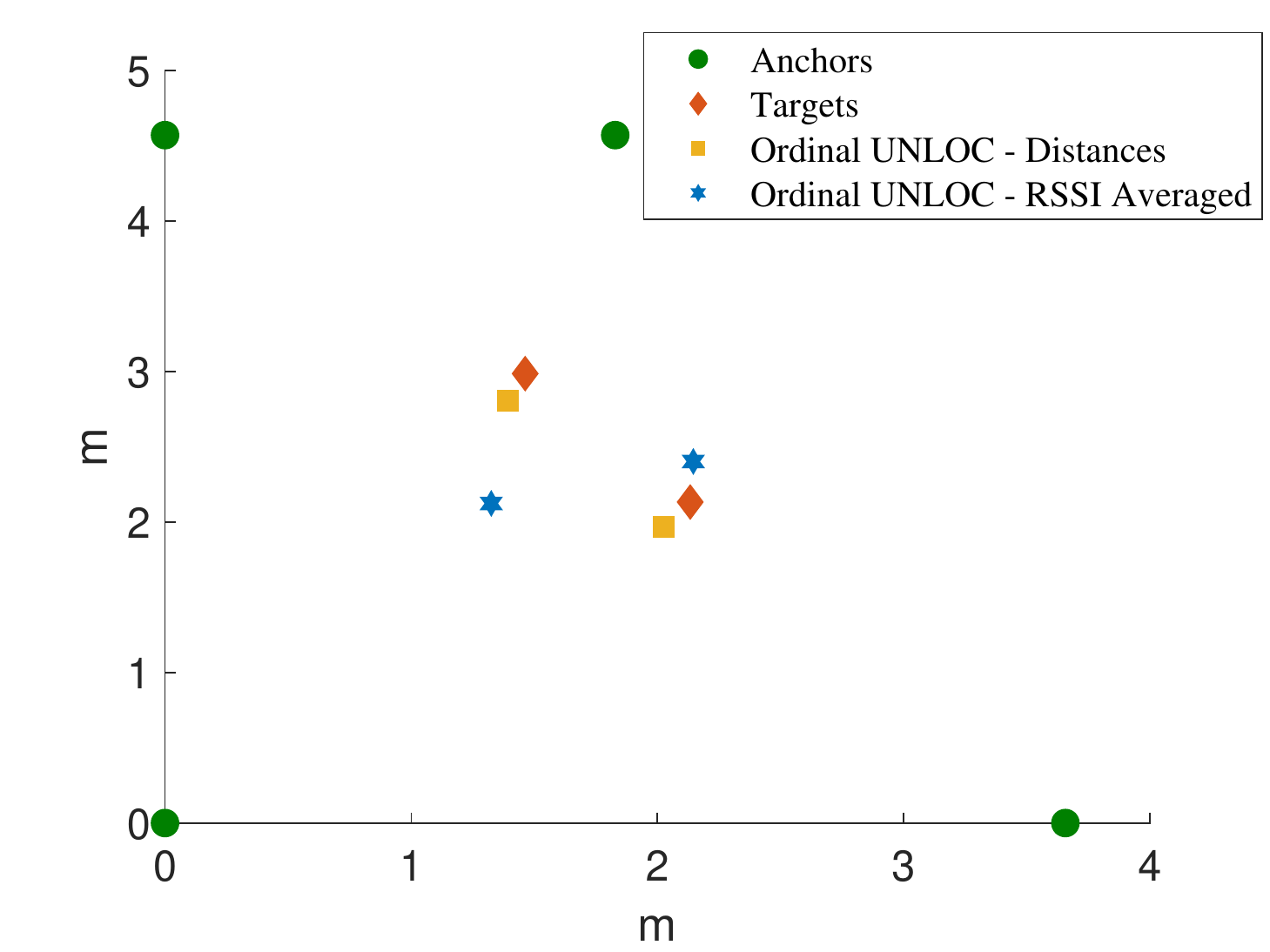}
	\caption{Hardware experiment to validate Ordinal UNLOC. Four anchors are placed  at  the  approximate  corners  of  a  rectangle  and  of many experiments conducted, two  targets  placed within  the  anchor  space are shown here. Here, both of the targets are localized simultaneously; the targets are individually localized in Figure \ref{Fig:config1_m} and Figure \ref{Fig:config2_m}. Ordinal  UNLOC  is  applied  on  both the  actual  cross-distances  and  the RSSI  values.  Ordinal UNLOC with no calibration outperforms UNLOC  with  calibrated  distance  estimates.  Averaging  all  Ordinal  UNLOC estimates with individual RSSI values provides an accurate location estimate, validating the Ordinal UNLOC algorithm.
	} 
	\label{Fig:config3_m}
\end{figure}

Two target nodes are shown below as examples. Using these two targets, we consider three localization configurations. In the first two configurations, the targets are localized individually (see Figure \ref{Fig:config1_m} and Figure \ref{Fig:config2_m}). In the third configuration, both targets are localized simultaneously (see Figure \ref{Fig:config3_m}). In each of the figures, the anchors are represented by green circles, and the targets locations by the red diamonds. 

Four scenarios were used for estimating the location of the target. In the first idealized scenario, Ordinal UNLOC was used to estimate the location of the target, when the ground-truth cross-distances between the target and the anchors and the actual distances between the anchors is assumed to be known to the algorithm. The second estimation scenario uses the RSSI values obtained from the experiments directly. The procedure followed was as described above. 

The final scenario was conducted to provide insight into the effectiveness of the Ordinal UNLOC algorithm in highly reflective environments. In this case, we used (\ref{eqn:receive_power}) to estimate the distances between the target and the anchors, with a fixed $G$ of 4, estimated using the techniques in \cite{banavar2018}. With these distance estimates, the UNLOC algorithm, described in Section \ref{ssec:UNLOC} and \cite{Sun2018}, is used to estimate the target location. We compute estimates for all RSSI values, and only show \textit{the best estimate}. 

Note that when multiple targets are simultaneously localized (see Figure \ref{Fig:config3_m}), we do not compare against the UNLOC algorithm, since UNLOC is designed to localize one target at a time. Furthermore, to keep the figure from being cluttered, we only show the averaged localization estimation when using RSSI values with Ordinal UNLOC, and omit showing the individual estimates. 

As can be seen from the results, the best performer is the case when the actual distances are known. While this is not possible in a real localization problem, it is used here to establish a performance benchmark. Using RSSI values directly with ordinal UNLOC provides better performance in spite of not needing to calibrate, and outperforms the (standard) UNLOC algorithm using calibrated values for the operating environment. The location estimate obtained by averaging all location estimates obtained by ordinal UNLOC on RSSI values provides the best performance of all the realistic scenarios. 

The Ordinal UNLOC algorithm was successful in estimating the location of the target devices in three different configurations and four different scenarios, in a rich scattering environment, outperforming conventional localization methods. Recall that we had over 200,000 samples of data recorded for our experiments, from which we calculated over 2000 location estimates from twenty locations of targets. From this data, we estimate that the normalized error when using Ordinal UNLOC is on the order of 0.02$m/m^{2}$. This is comparable to, but a little worse than the normalized error expected from using carefully calibrated fingerprinting methods (normally between 0.01$m/m^{2}$ and 0.001$m/m^{2}$) \cite{Khandker2020}. However, we should note here that fingerprinting methods require significant precomputing and training, and are reliable only when the environment does not change. This, in addition to our simulation results in Sections \ref{ssec:benchmark_tests} and \ref{ssec:communications_models}, validate the effectiveness of our approach. 

\subsection{Latency}
\label{ssec:latency}

To implement Ordinal UNLOC successfully, we require one sample for good localization. Based on our hardware experiments, we see that one in ten samples provides an accurate estimate. That is, the latency of our approach is on the order of 10 measurements per location estimate. In fingerprinting work such as \cite{faragher15, Zhuang2016, jin15, Dani2017, minho_database2017, Khandker2020, chen2019learning, chen2019wifi, Akino2020, yoo2020fingerprint, zheng2020deep, he2020hybrid, al2020comparative, zhu2020indoor}, on the order of 10-100 samples are needed for each estimate, showing that our work is competitive when it comes to latency. Note that although there are advantages to using fingerprinting methods for localization, these methods require extensive mapping of the environment, \textit{a priori}, leading to challenges in dynamic environments, as well as maintaining a sufficiently large database to provide robustness to the algorithm.

\section{Conclusions and Future Work}
\label{sec:conclusions}

In this paper, we present a new algorithm for performing localization with limited and noisy measurement data between anchors and targets. Since estimating distances from this limited data is typically unreliable, we instead use the measurements to compute ordinal comparisons for the distances between all the anchors and targets. We used the ordinal data and designed a three-step ``Ordinal UNLOC'' algorithm for localization: (1) rank aggregation; (2) function learning; and (3) unfolding localization. Through simulation results, we demonstrate the effectiveness of our algorithm. Hardware experiments, where Ordinal UNLOC is used to localize devices using Bluetooth RSSI, were successful, suggesting that the algorithm can be used in real settings, including rich scattering environments. Furthermore, unlike traditional methods which require calibration depending on the operating environment, Ordinal UNLOC can be directly applied to measured distance proxies such as TOA or RSSI. 

While we have demonstrated the effectiveness of the approach, we also note a couple of limitations: the algorithm consists of a three stage process which can become computationally expensive, and for the $\mathbb{Z}$ tensor to be constructed, we require distance or distance proxy cross-information between all nodes, anchors and targets.

Future work involves the investigation of the role of anchor locations on localization. While ordinal data is one-bit data, another avenue of exploration is the effect of increasing the number of bits available for approximating incomplete and noisy measurements. We will investigate further, the complexity and latency of Ordinal UNLOC and compare it with other methods such as traditional multilateration and fingerprinting.

\bibliographystyle{IEEEtran}
\bibliography{sensors}

\vfill

\end{document}